%% file: hnunu_paper.tex
\RequirePackage{lineno}
\documentclass[aps,prl,twocolumn,superscriptaddress,showpacs,preprintnumbers]{revtex4}

\usepackage{tcolorbox}
\usepackage{ulem} 
\usepackage{graphicx} 
\usepackage{dcolumn}  
\usepackage{xspace}
\usepackage{hepunit}
\usepackage{hyperref}
\usepackage[capitalise]{cleveref} 
\usepackage{calc}
\usepackage{amsmath, amssymb, amsfonts, mathrsfs, amsbsy}
\usepackage{xstring}
\usepackage{units}
\usepackage{booktabs}
\usepackage[compatibility=false,format=plain]{caption}
\usepackage[subrefformat=parens,labelformat=parens]{subcaption}
\usepackage{lineno}
\include{variable_definitions}
\include{owncommands}

\graphicspath{{ps}}

\newcolumntype{C}{>{$}c<{$}}
\AtBeginDocument{
\heavyrulewidth=.08em
\lightrulewidth=.05em
\cmidrulewidth=.03em
\belowrulesep=.65ex
\belowbottomsep=0pt
\aboverulesep=.4ex
\abovetopsep=0pt
\cmidrulesep=\doublerulesep
\cmidrulekern=.5em
\defaultaddspace=.5em
}

\begin{document}

\title{
\begin{flushright}                           
{\small Belle Preprint 2017-10}\\
{\small KEK Preprint 2017-6}\\
\end{flushright} 
\quad\\[1.0cm]
Search for $\boldsymbol{\boldchan{0}}$ decays with semileptonic tagging at Belle}
\input{author}

\noaffiliation

\begin{abstract}
  We present the results of a search for the rare decays \mbox{\chan{0}\!\!,} where \textit{h} stands for $K^+,\:\ks,\:\kstarP,\:\kstarO,\:\pi^+,\:\pi^0,\:\rho^+$ and $\rho^{0}$. The results are obtained with $772\times10^{6}$ $B\overline{B}$ pairs collected with the Belle detector at the KEKB $e^+ e^-$ collider. We reconstruct one $B$ meson in a semileptonic decay and require a single $h$ meson but nothing else on the signal side. We observe no significant signal and set upper limits on the branching fractions. The limits set on the \mbox{\chan{310}\!\!,} \chan{313}\!\!, \chan{211}\!\!, \chan{111}\!\!, \chan{213}\!\!, and \chan{113} channels are the world's most stringent.
\end{abstract}


\maketitle

\tighten

{\renewcommand{\thefootnote}{\fnsymbol{footnote}}}
\setcounter{footnote}{0}

\begin{figure}
  \centering
  \begin{subfigure}[b]{0.49\linewidth}
    \includegraphics[width=\linewidth]{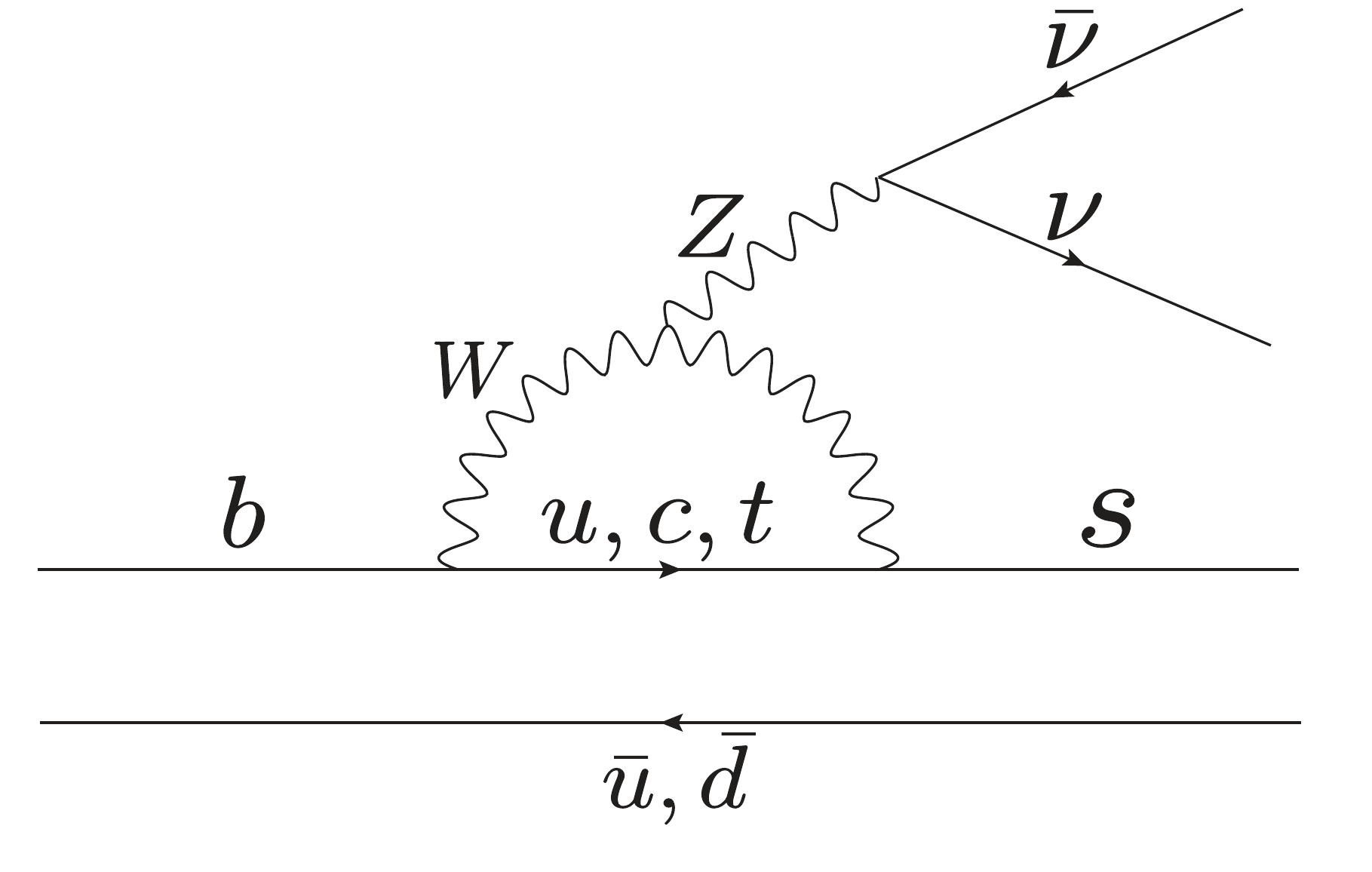}
    \caption{penguin}
  \end{subfigure}
  \begin{subfigure}[b]{0.49\linewidth}
    \includegraphics[width=\linewidth]{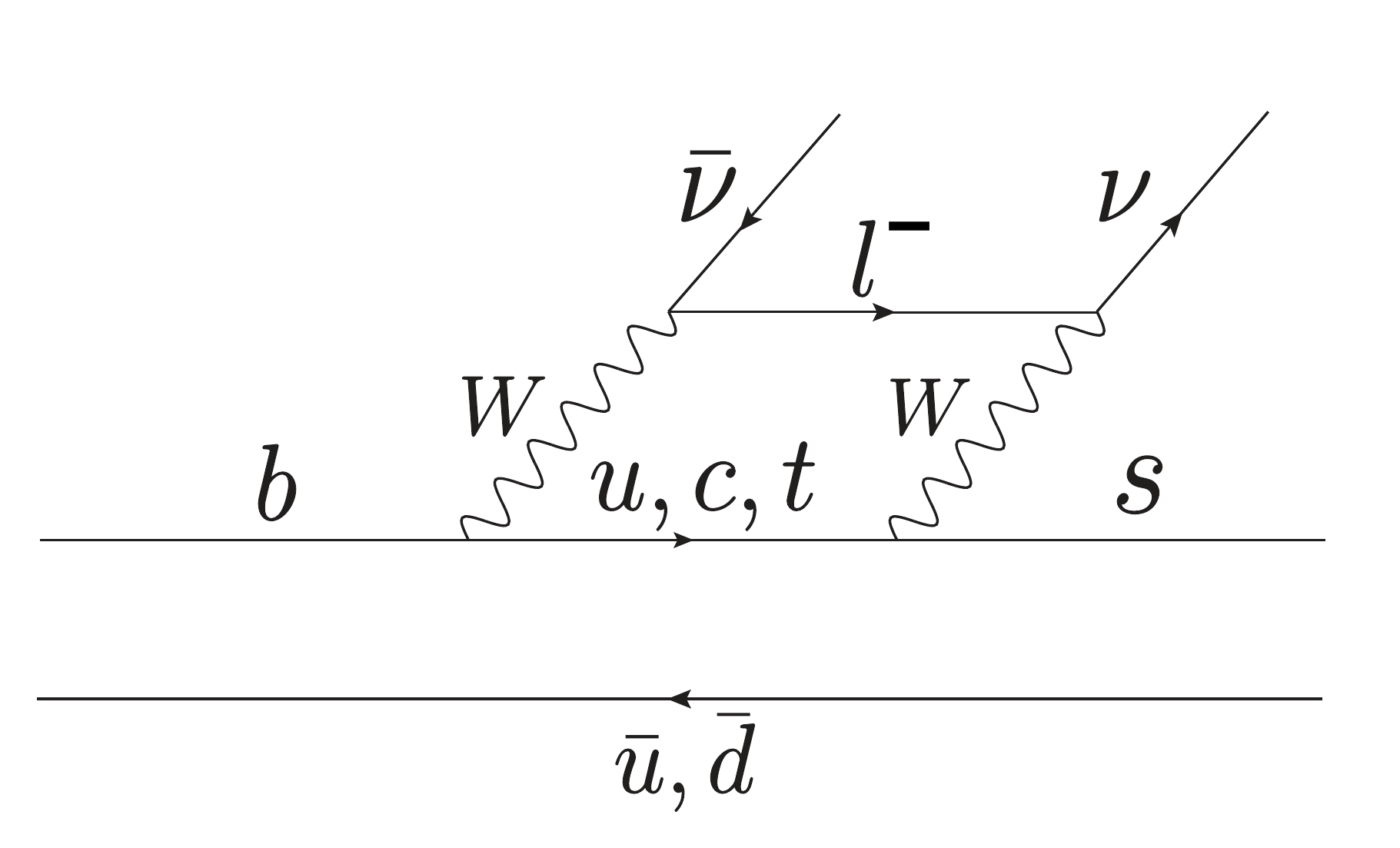}
    \caption{box}
  \end{subfigure}
  \caption{Lowest-order SM quark level diagrams for the \protect\chan{321} channel; the diagrams for other channels are analogous.}
  \label{fig:feynman}
\end{figure}

The decays \chan{0}~\cite{charge_conj} can proceed only via a penguin or a box diagram at leading order in the standard model (SM), as shown in  \cref{fig:feynman}, and are thus highly suppressed~\cite{Buras_knunu_14}. Theoretical calculations for the branching fractions cover the range from $1.2\times10^{-7}$~\cite{pinunu-prediction} \mbox{(\!\!\chan{111}\!\!)} to $9.2\times10^{-6}$~\cite{Buras_knunu_14} (\!\!\chan{323}\!\!). Recent results by LHCb~\cite{LHCB_Kll_15, LHCb_RK} show evidence for a deviation of experimental data from expected values in the angular observable $P_5^\prime$ in $B^0\to K^{\ast0} \mu^+\mu^-$ decays, and in the ratio of the $B^+\to K^+ \mu^+\mu^-$ to $B^+\to K^+ e^+e^-$ branching fractions. A measurement of $P_5^\prime$ by Belle~\cite{belle_kll} is compatible with both, the SM prediction and the LHCb result. Different new physics models proposed to explain these observations can also influence $B\to K^{(\ast)}\nu\overline{\nu}$ decays. Therefore, \chan{0} channels provide an important test for any model proposed to solve these tensions. Additionally, \chan{0} channels are theoretically clean due to the mediation of the transition by a $Z$ boson alone, in contrast to $B\to K^{(\ast)} l^+l^-$ decays~\cite{Buras_knunu_14} where the photon contributes.

\chan{0} decays have been studied previously by Belle with a hadronic tagging algorithm~\cite{Belle_hadronic_13}, and by BaBar utilizing both hadronic~\cite{Babar_hadronic_13} and semileptonic tagging~\cite{babar_semileptonic_10}. Recent results by Belle~\cite{semileptonic-b-to-dstar-tau} have shown that the usage of semileptonic tagging enhances the sensitivity of some analyses significantly.~The semileptonically tagged sample provides a statistically independent and more efficiently tagged data set of reconstructed $B\overline{B}$ events as compared to the hadronically tagged sample. 

We search for \chan{0} decays with the full Belle data sample produced by the KEKB collider~\cite{kekb_accelerator} at the \YfourS center-of-mass (CM) energy with an integrated luminosity of $\unit[711]{fb^{-1}}$, corresponding to $(772\pm 11)\times 10^{6}$ $B\overline{B}$ pairs. A data set of $\unit[89]{fb^{-1}}$ taken at an energy $\unit[60]{MeV/c^2}$ below the resonance energy is used to study background from $e^+e^-\to q\overline{q}$ processes (continuum), where $q\in u, d, s, c$. We refer to this data set as the off-resonance sample. We model the decays with the \texttt{EVTGEN} package~\cite{evtgen} and simulate the detector response with the~\texttt{GEANT3} package~\cite{geant}. We include a randomly-triggered sample to account for beam-related background. The signal process is modeled according to three-body phase space.

The Belle detector~\cite{belle_tdr} is a large-solid-angle magnetic
spectrometer that consists of a silicon vertex detector (SVD),
a 50-layer central drift chamber (CDC), an array of
aerogel threshold Cherenkov counters (ACC),
a barrel-like arrangement of time-of-flight
scintillation counters (TOF), and an electromagnetic calorimeter
comprised of CsI(Tl) crystals (ECL) located inside
a super-conducting solenoid coil that provides a 1.5~T
magnetic field.  An iron flux-return located outside of
the coil is instrumented to detect \klong\:mesons and to identify
muons (KLM). The detector
is described in detail elsewhere~\cite{belle_tdr}.
Two inner detector configurations were used. A 2.0~cm beampipe
and a 3-layer silicon vertex detector were used for the first sample
of $152 \times 10^6 B\overline{B}$ pairs, while a 1.5~cm beampipe, a 4-layer
silicon detector and a small-cell inner drift chamber were used to record
the remaining $620 \times 10^6 B\overline{B}$ pairs~\cite{svd2}.

The three-body \chan{0} decay, with two invisible particles in the final state, does not convey sufficient kinematic information to isolate the signal. Thus, we first reconstruct the accompanying $B$ meson $\left(\btag\!\right)$ in the semileptonic decay channels $B\to D^{(\ast)}l\nu_l$ ($l=e,\mu$), where neutral (charged) $D$ candidates are reconstructed in 10 (7) different decay channels. This amounts to 108 different decay channels. The tagging algorithm, described elsewhere~\cite{FullReco_paper,B_tau_nu_semilep}, uses multiple instances of neural network classifiers built using the NeuroBayes package~\cite{Neurbayes_paper} in a hierarchical approach to find \btag candidates. The output of the neural network used to identify real \btag candidates transformed into the interval $\left[0,1\right]$ is referred to as \tagout\!\! and can be interpreted as the probability of the \btag meson to be a true $B$ in a generic sample. We combine \btag candidates with our signal selection to form signal event candidates. We separate charged pion and kaon candidates based on particle identification (PID) selection criteria utilizing CDC, ACC and TOF information. We combine the PID information in a likelihood ratio $\mathcal{P}_{K\pi}=\mathcal{L}_K/\left(\mathcal{L}_K+\mathcal{L}_{\pi}\right)$, where $\mathcal{P}_{K\pi}$ is a function of the polar angle and the momentum of the track in the laboratory system. We require $\mathcal{P}_{K\pi} > 0.6$ $(<0.4)$ for $K^\pm$ $(\pi^\pm)$ candidates. The kaon (pion) identification efficiency is $88\,\%-93\,\%$ $(86\,\%-93\,\%)$ with a $\pi$ $(K)$ misidentification probability of $10\,\%-12\,\%$ $(8\,\%-11\,\%)$. 

We reduce the number of poor quality tracks by requiring that  $dz\left(dr\right) < \unit[4\left(2\right)]{cm}$, where $dz$ \mbox{($dr$)} is the distances of closest approach of a track to the interaction point along \mbox{(transverse to)} the $z$ axis, which is antiparallel to the positron beam. Signal $B$ daughter candidates are reconstructed through the decays $K^{*0}\to K^+\pi^-$, $K^{*+}\to K^+\pi^0$ and $\ks\pi^+$, $\rho^+\to\pi^+\pi^0$, $\rho^0\to \pi^+\pi^-$, $\ks\to \pi^+\pi^-$, and $\pi^0\to \gamma\gamma$. \ks candidates are selected following Ref.~\cite{goodKs}. Photons used for $\pi^0$ reconstruction are required to have a minimal energy of $\unit[50]{MeV/c^2},\:\unit[100]{MeV/c^2},\:\unit[150]{MeV/c^2}$ for the barrel ($\theta\in \left[32^\circ, 129^\circ\right]$), forward ($\left[17^\circ, 32^\circ\right]$), and backward ($\left[129^\circ,150^\circ\right]$) region of the ECL, respectively, where $\theta$ is taken with respect to the $z$ axis. The invariant mass of the two $\gamma$ candidates is required to fulfill $M_{\gamma\gamma}\in \left[118, 150\right]\unit{MeV/c^2}$, while the invariant mass of the $K^\ast$ ($\rho$) candidates is required to be within $\unit[150]{MeV/c^2}$ $(\unit[250]{MeV/c^2})$ of the nominal mass from Ref.~\cite{PDG_booklet}. The mass requirements are subsequently optimized using Monte Carlo (MC) simulations by maximizing the figure of merit $N_R/\sqrt{N_R+N_F}$, where $N_R$ is the number of correctly reconstructed mesons and $N_F$ the number of fake candidates, both passing the requirement. We combine a \btag candidate with the reconstructed signal-$B$ decay product (\hsig) to form an \YfourS candidate.

Events with additional charged tracks or $\pi^0$ candidates that satisfy our selection criteria are rejected. Furthermore, we remove events with two or more tracks not fulfilling our requirement on $dr$ or $dz$ ``raw tracks.'' We veto events with reconstructed \klong\: candidates and weight our background simulations to account for known data--MC differences, as described in Ref.~\cite{Hara_btaunu}. An important variable to identify correctly--reconstructed signal events is the extra energy, \EEcl\!\!. We sum all ECL clusters not used in the reconstruction of the \YfourS candidate, not associated with a track, and fulfilling the same energy requirements as the clusters used to form $\pi^0$ candidates. We require $\EEcl < \unit[1.2]{GeV/c^2}$. We also require the momentum of the \hsig\: candidate in the CM system to fulfill $\pcms\!\in \left[0.5, 2.96\right]\,\mathrm{GeV}/\mathrm{c}^2$, the missing energy in the CM system $\emiss > \unit[2.5]{GeV/c^2}$, the momentum of the \btag lepton candidate in the CM system $\taglpcms < \unit[2.5]{GeV/c^2}$, and a minimal tag quality of $\tagout > 0.005$. These requirements are motivated by kinematic boundaries, data--MC differences in case of low--momentum \hsig\!, and badly reconstructed tag candidates. To suppress pions from $D$ decays misreconstructed as muons, we veto events where the invariant mass of the $K$ (\kstarO) candidate and the tag--side lepton fulfill $M_{Kl}\in\left[1.85, 1.87\right]\unit[]{GeV/c^2}$. The channel--dependent fraction of events with more than one \YfourS candidate can be as large as $\unit[20]{\%}$, dominated by candidate exchange between signal-- and tag--side. In such cases, we select the candidate with the highest \tagout value, \textit{i.e.}\!, the candidate with the highest probability of being correctly reconstructed. In MC studies, we find that the efficiency of this selection is between $\unit[65]{\%}$ \mbox{(\!\!\chan{213}\!\!)} and $\unit[92]{\%}$ \mbox{(\!\!\chan{310}\!\!)}.

We reconstruct tagged $B^+\to\overline{D^0}\left(K^+\pi^-\right)\pi^+$ and $B^0\to D^-\left(K^+\pi^-\pi^-\right)\pi^+$ decays to correct for experimental data--MC efficiency differences. Both channels can be reconstructed with negligible background and are well described in MC. We bin \tagout equally in 4 (3) bins for charged (neutral) $B$ mesons and calculate the number of reconstructed events in data and MC. We assign the ratio as a weight in each bin of \tagout\!\!. This calibration includes a correction of the tagging efficiency $\times$ the number of $B\overline{B}$ pairs produced ($N_{B\overline{B}}$) $\times$ the branching fraction of $\YfourS$ to charged and neutral $B$ meson pairs, as we have a separate calibration for $B^+$ and $B^0$. 
We train one neural network per channel to suppress continuum events. We use 16 modified Fox-Wolfram moments~\cite{ksfw}, nine CLEO cones~\cite{cleo_cones_paper}, the cosine of the angle of the thrust axis relative to the $z$ axis, and the angle of the momentum of the \btag candidate with respect to the $z$ axis. We refer to the output of this neural network as \csout\!\!.

To optimally separate signal from background, another neural network is trained for each reconstructed channel. We optimize the requirement on the network output \mbox{(\finalout)} by maximizing a figure-of-merit $\varepsilon / (\frac{n_\sigma}{2}+\sqrt{N_B})$, which is independent of the signal-to-background ratio and optimized for searches~\cite{Punzi_fom}. Here, $\varepsilon$ is the signal efficiency while $N_B$ denotes the number of background events passing the requirement on \finalout\!\!. Both values are determined from MC. We choose a desired significance $n_\sigma = 3$. The most powerful variables to identify the signal are \pcms\!, \csout\!, the cosine between the momentum of the $D^{(\ast)}l$ system and the momentum of the \btag in the CM system~\cite{ctbdl}, the cosine of the angle of the missing momentum relative to the $z$ axis, the cosine of the angle of the thrust axis, \taglpcms\!\!, and, for the $\rho$ and $K^\ast$ channels, the reconstructed invariant mass. The number of input variables varies for each channel, spanning a range from 17 to 31. 

We evaluate the description of the data by our MC by looking into an \EEcl sideband \mbox{($\EEcl>\unit[0.3]{GeV/c^2}$)}, by reconstructing tagged $B\to D^\ast l\nu_l$ decays, and by utilizing the off-resonance sample. We find good agreement between data and MC in the \EEcl sidebands for six of the eight channels. However, we find an underestimation of continuum background in MC in the \chan{321} and  the  \chan{211} channels, which we correct by scaling the continuum component in the background model by the observed data--MC ratio in the off-resonance sample.

To extract the signal yield in each channel, we perform an extended binned maximum likelihood (ML) fit to the \EEcl distribution. We use histogram templates to model signal as well as backgrounds from charm $B$-decay ($b\to c$), charmless $B$-decay ($b\to s, u, d$), and continuum. We fix the relative fractions of the background components to MC expectations and leave only the signal and the overall background yields as freely floating parameters. We perform extensive toy MC studies to estimate the sensitivity of our procedure. For this purpose, we simulate 1000 background-only samples for each channel and calculate an expected limit on the signal yield by integrating the profile likelihood up to the point where it includes $\unit[90]{\%}$ of the positive region. We also simulate samples with various numbers of signal events to test for a possible bias. We find a non-negligible but modest bias in almost all investigated channels. We fit this bias with a linear function, whose slope is consistent with 1.0 and whose intercept lies between $0$ and $-2$ events. We correct for this bias in our fit to data. 

\begin{figure}[h]
  \centering
  \begin{subfigure}{0.49\linewidth}
    \includegraphics[width=0.7\linewidth,angle=270]{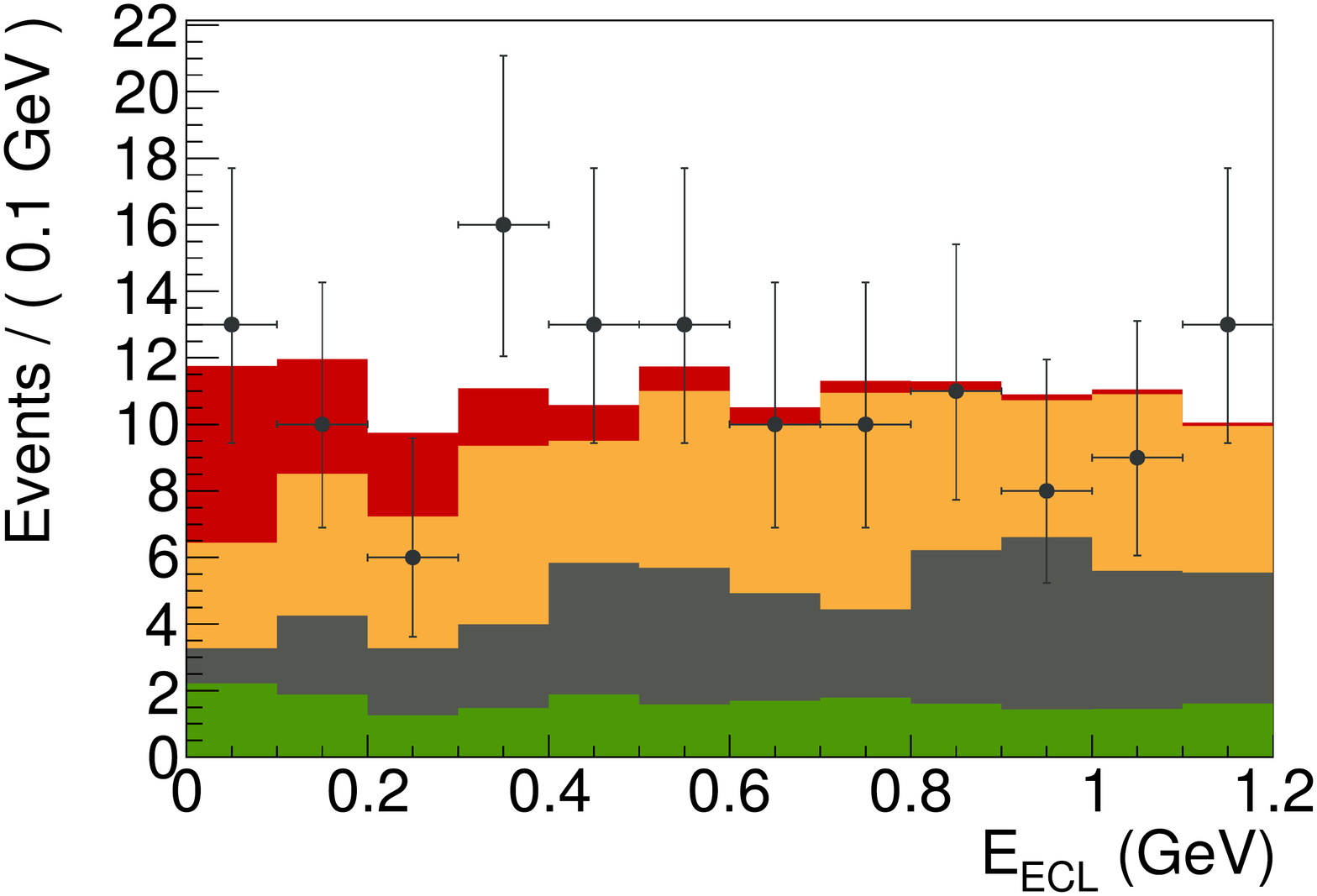}
    \caption{\protect\chan{321}}
  \end{subfigure}
  \begin{subfigure}{0.49\linewidth}
    \includegraphics[width=0.7\linewidth,angle=270]{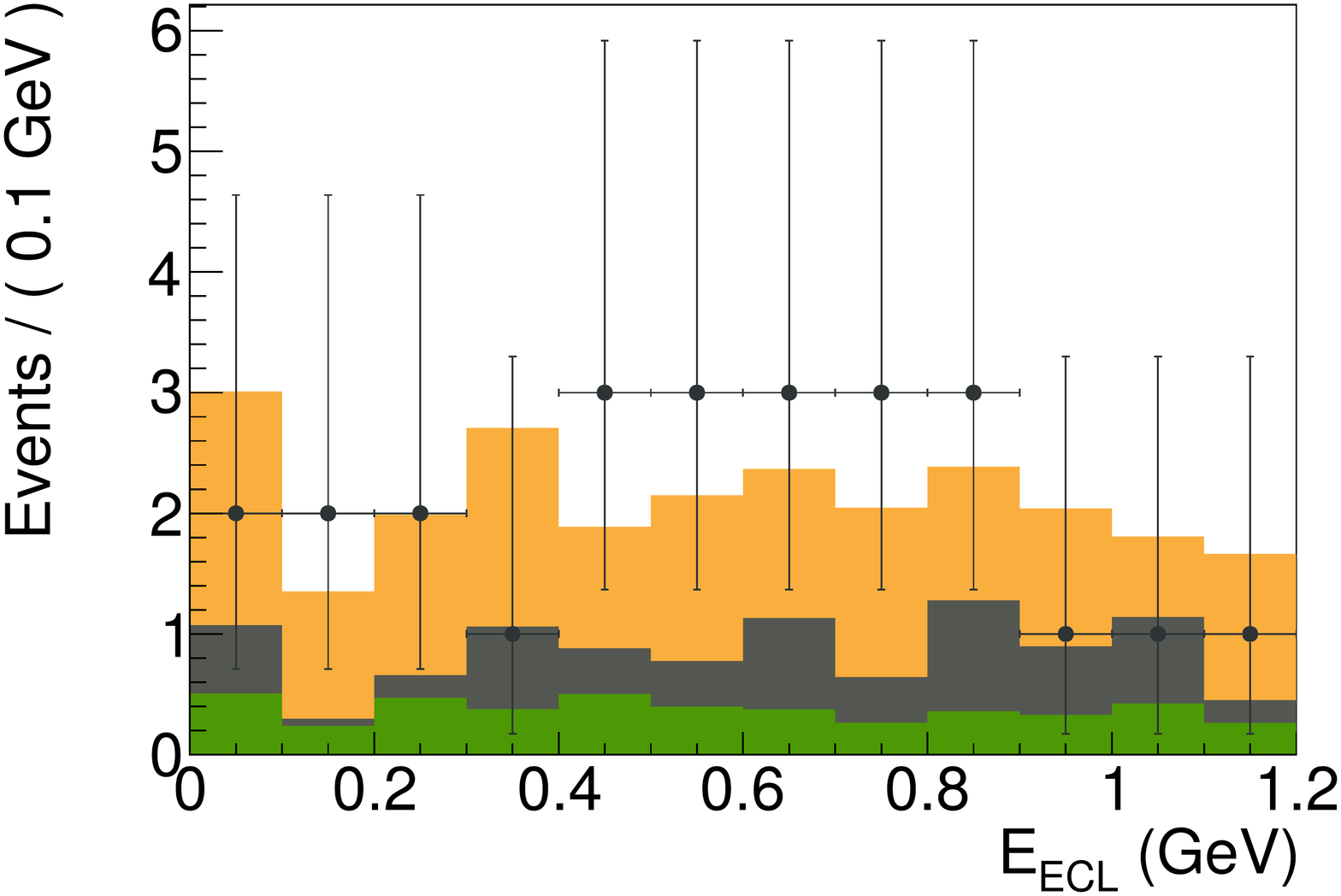}
    \caption{\protect\chan{310}}
  \end{subfigure}
  \begin{subfigure}{0.49\linewidth}
    \includegraphics[width=0.7\linewidth,angle=270]{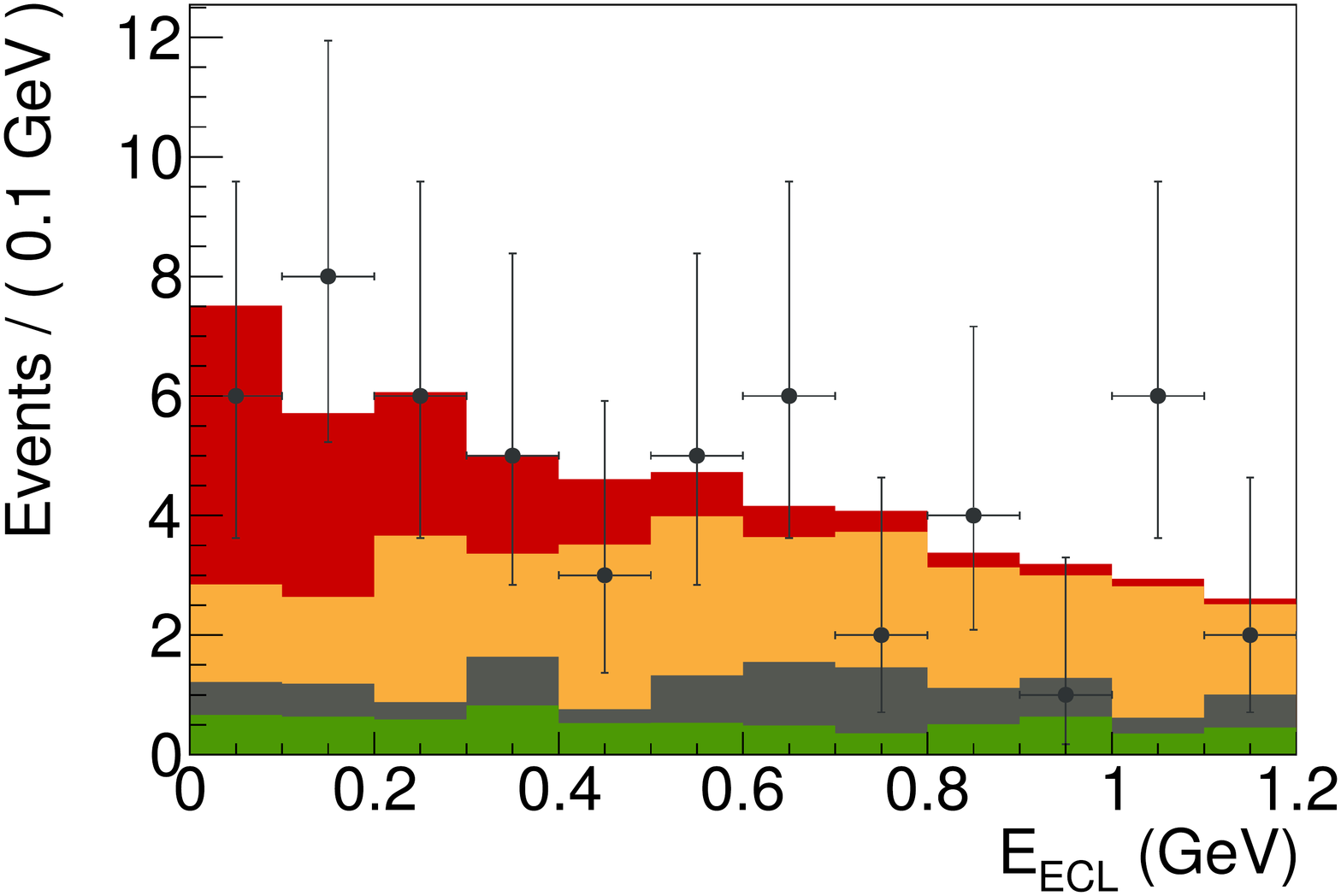}
    \caption{\protect\chan{323}}
  \end{subfigure}
  \begin{subfigure}{0.49\linewidth}
    \includegraphics[width=0.7\linewidth,angle=270]{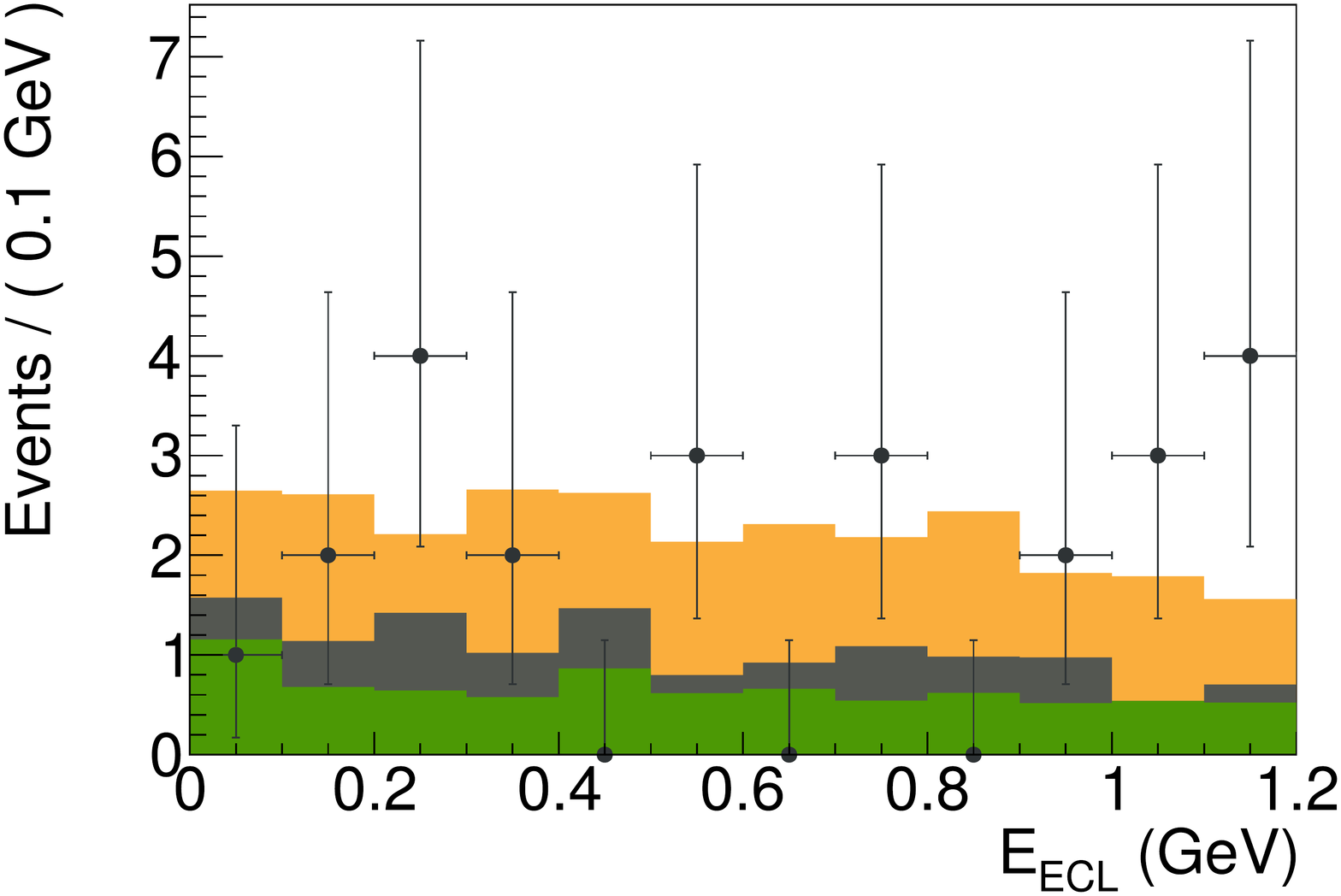}
    \caption{\protect\chan{313}}
  \end{subfigure}
  \begin{subfigure}{0.49\linewidth}
    \includegraphics[width=0.7\linewidth,angle=270]{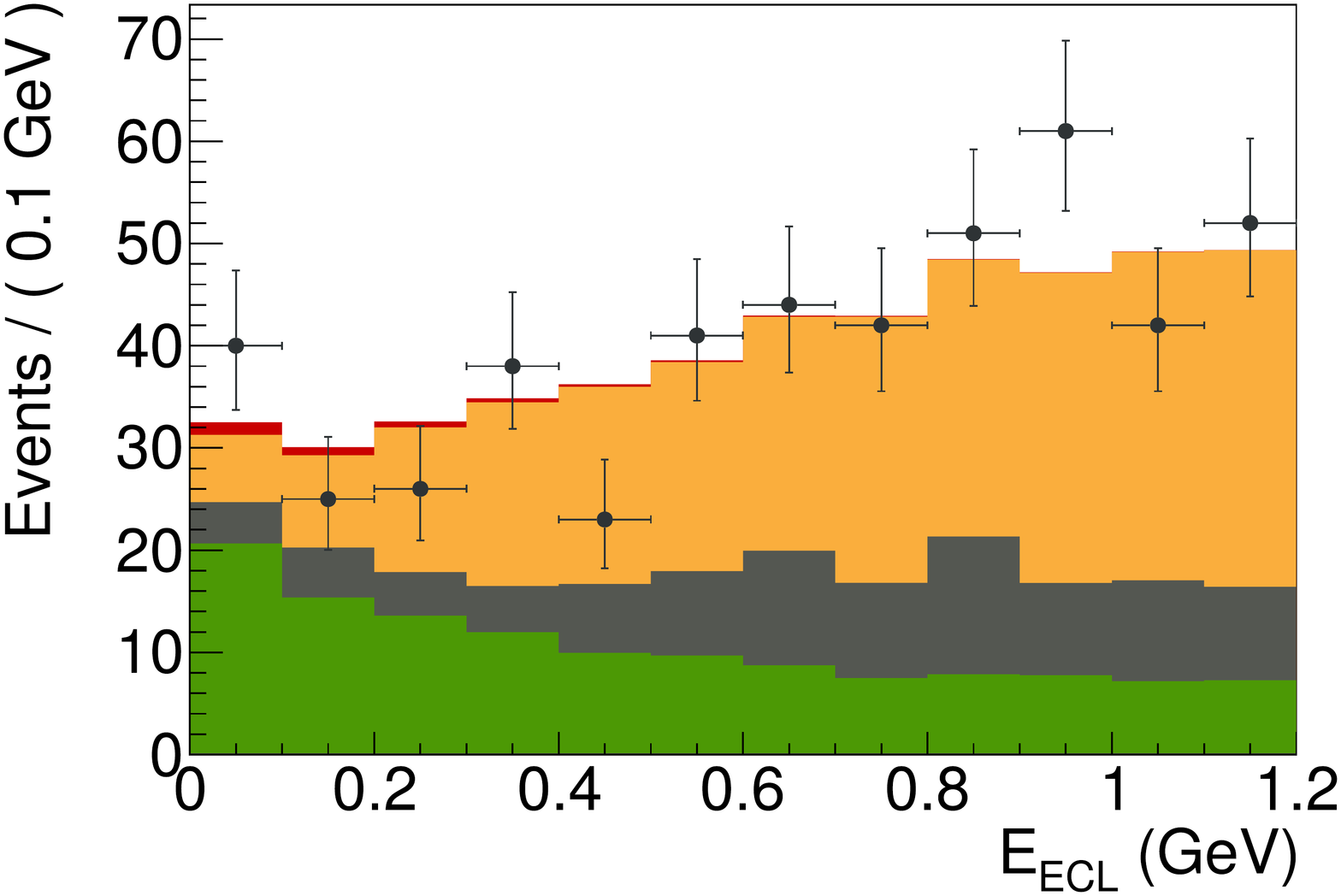}
    \caption{\protect\chan{211}}
  \end{subfigure}
  \begin{subfigure}{0.49\linewidth}
    \includegraphics[width=0.7\linewidth,angle=270]{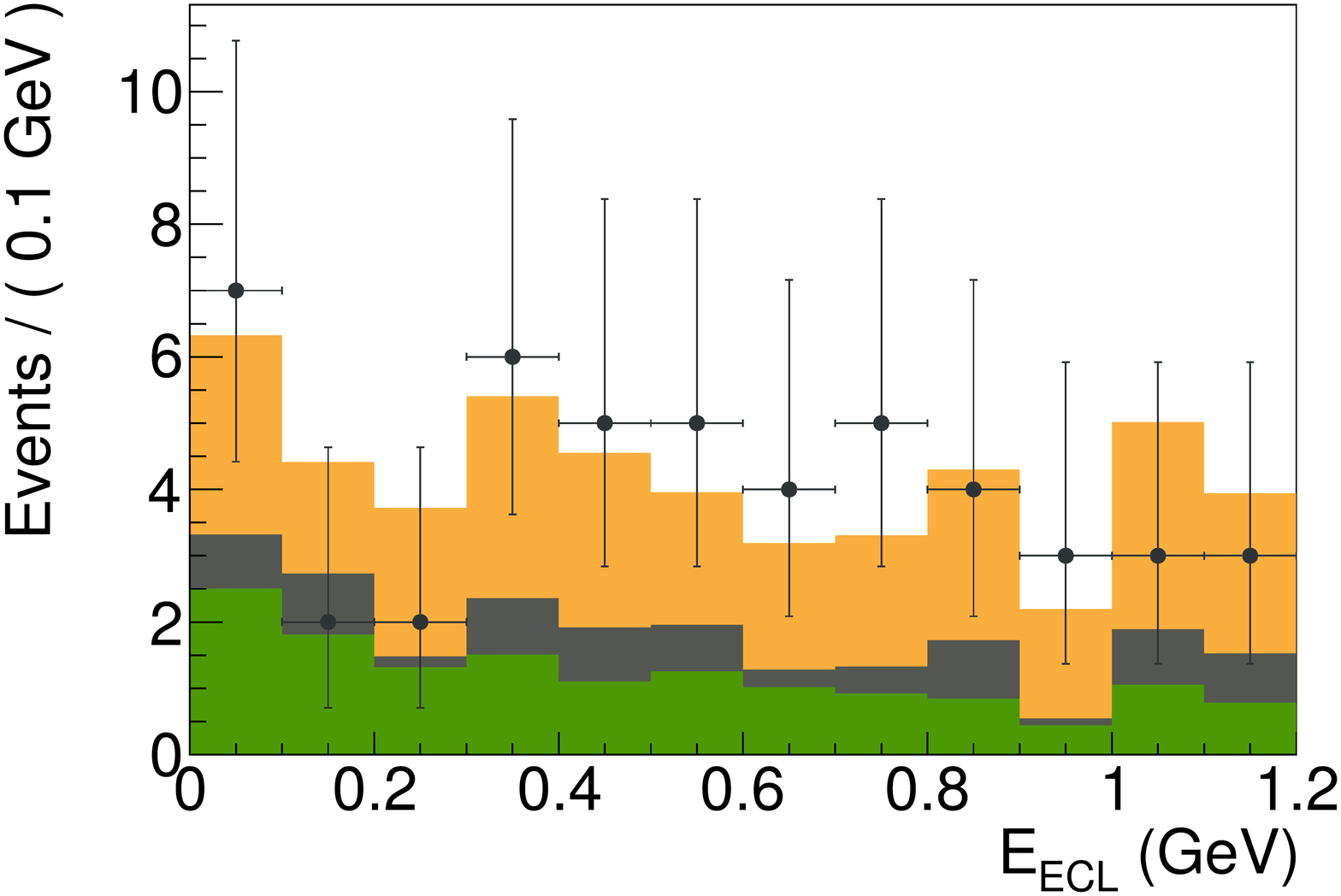}
    \caption{\protect\chan{111}}
  \end{subfigure}
  \begin{subfigure}{0.49\linewidth}
    \includegraphics[width=0.7\linewidth,angle=270]{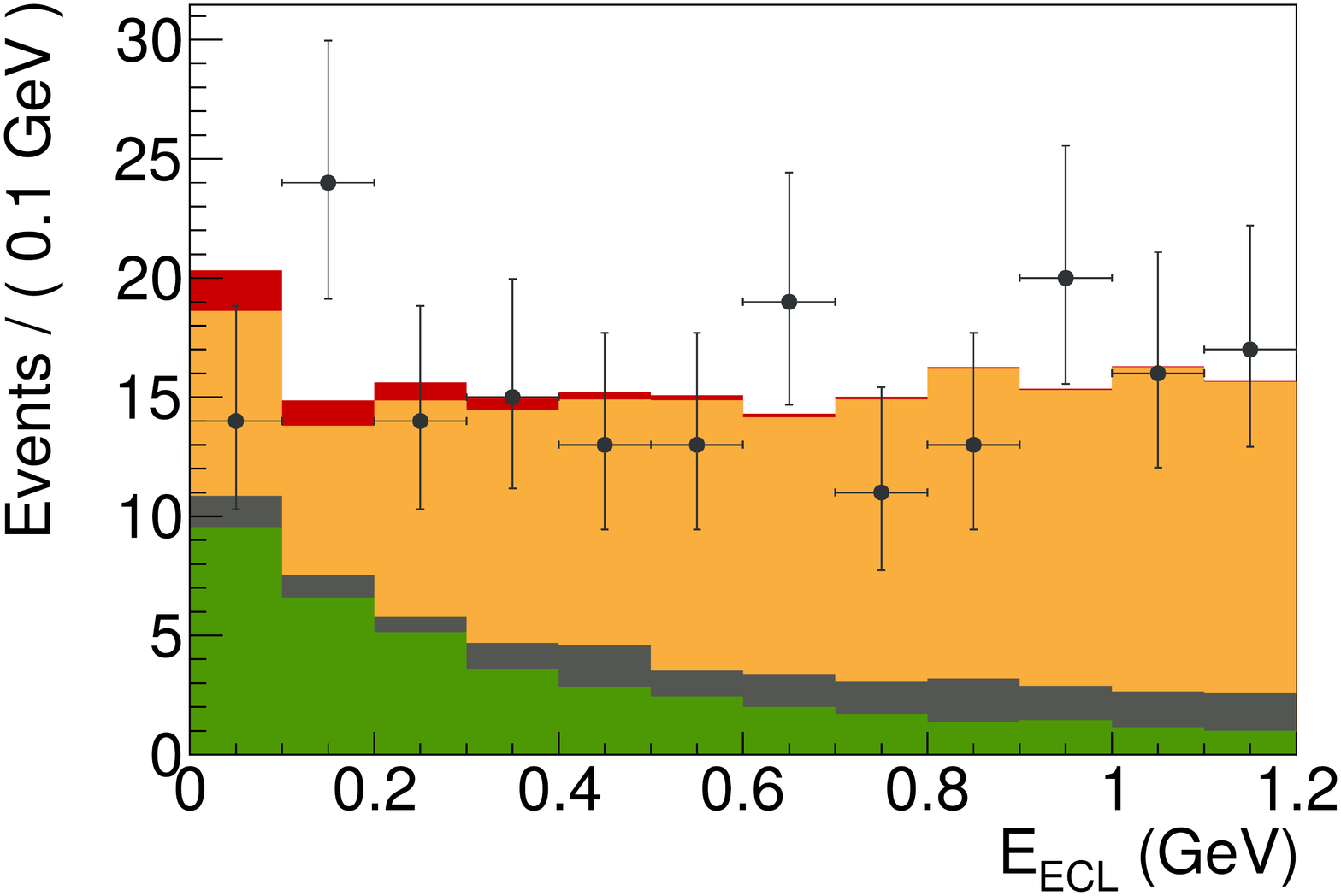}
    \caption{\protect\chan{213}}
  \end{subfigure}
  \begin{subfigure}{0.49\linewidth}
    \includegraphics[width=0.7\linewidth,angle=270]{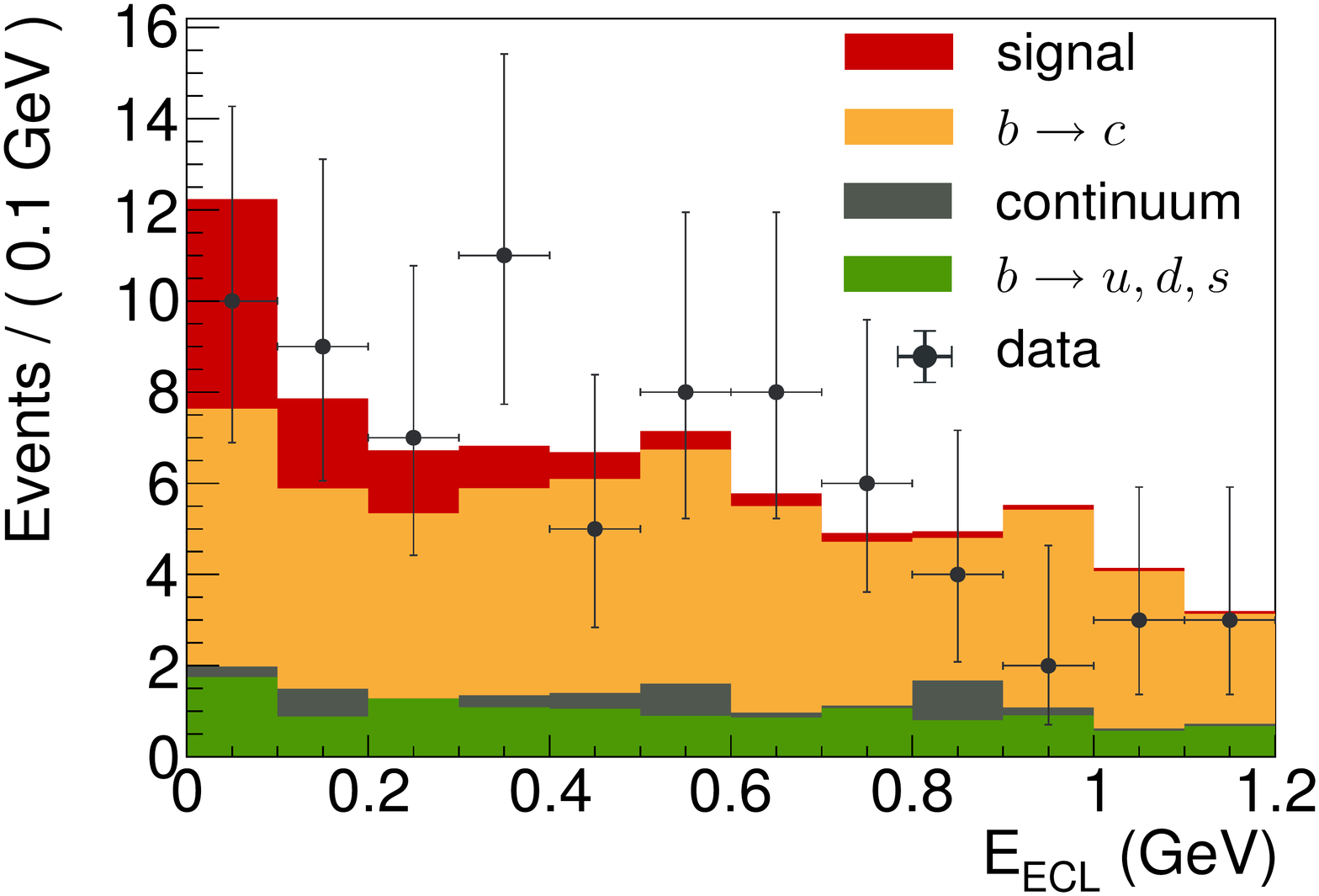}
    \caption{\protect\chan{113}}
  \end{subfigure}
  \caption{\EEcl distributions for all eight \protect\chan{0} channels.}
  \label{fig:fits}
\end{figure}

The fit results are listed in \cref{tab:events}; \cref{fig:fits} shows the distributions of the data together with the fitted signal and background models. The fit yields no significant signal in any channel. The largest signal contribution is observed in the \chan{323} channel with a significance of $\unit[2.3]{\sigma}$. The significance is defined by evaluating the likelihood of the complete model $\mathcal{L}_{\mathrm{max}}$ and the background-only likelihood $\mathcal{L}_{0}$: \mbox{$S=\sqrt{2\log{\left(\mathcal{L}_{\mathrm{max}}/\mathcal{L}_{0}\right)}}$}. Both are evaluated at their respective best fitting point. We calculate the branching fraction of the $i$-th mode by \mbox{$\mathscr{B}^i=N^i_{\mathrm{sig}}/\left(\varepsilon^i_{\mathrm{rec}} \times N_{B\overline{B}}\right)$}, where the reconstruction efficiency $\varepsilon_{\mathrm{rec}}^{i}$ includes all daughter branching fractions. These efficiencies, along with the expected and measured $\unit[90]{\%}$ confidence level (C.L.) upper limit~\cite{CL} for each channel, are  displayed in \cref{tab:results}.

\begin{table}[h]
  \centering
  \caption{Results}
  \begin{subtable}{\linewidth}
    \caption{Observed signal yield (corrected for fitting bias) in each channel. The first error is statistical and the second is systematic.}
    \input{observed_events}
    \label{tab:events}
  \end{subtable}
  \label{tab:all_results}
  \begin{subtable}{\linewidth}
    \caption{Expected (median) and observed upper limits on the branching fraction at $\unit[90]{\%}$ C.L. The observed limits include the systematic uncertainties.}
    \input{measured_br_limits_result}
    \label{tab:results}
  \end{subtable}
\end{table}

\begin{figure}[h]
  \centering
  \includegraphics[width=\linewidth]{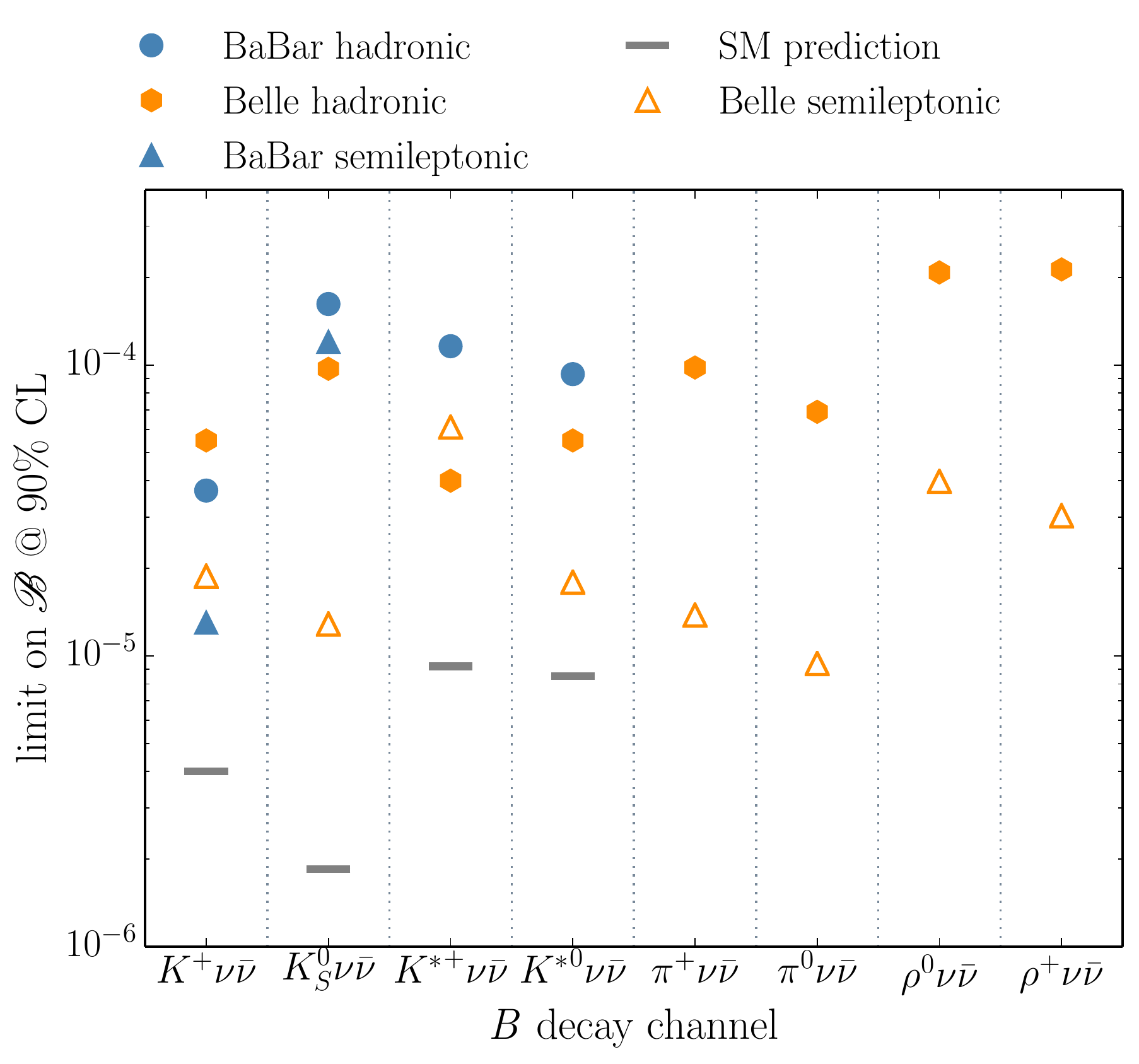}
  \caption{Observed limits for all channels in comparison to previous results for the BaBar measurement with semileptonic~\cite{babar_semileptonic_10} and hadronic tag~\cite{Babar_hadronic_13}, as well as the Belle measurement utilizing hadronic tagging~\cite{Belle_hadronic_13}. The theoretical predictions are taken from Ref.~\cite{Buras_knunu_14}.}
  \label{fig:final-plot}
\end{figure}

We estimate the uncertainty on the fixed fractions, the \klong\: veto efficiency, the continuum scaling, the tagging efficiency, and the fit bias correction by refitting the data with each of these quantities varied by $\pm 1\,\sigma$. We estimate the shape uncertainty by simulating 1000 toy templates obtained by drawing a random number from a Gaussian distribution with the mean and error of the respective bin of our fit model as the central value and deviation. The $\unit[\pm1]{\sigma}$ quantiles of the resulting distribution are used as estimators of the uncertainty. We estimate the uncertainty on the $\pi^0$ and charged track vetoes by comparing the respective efficiency differences between data and MC for the $B\to D\pi$ sample with and without the veto applied. We obtain a value of $4\,\%$ in both cases for charged and neutral channels alike. We evaluate the influence of the requirement on the number of raw tracks via the same sample by setting it to two and zero, respectively. We subsequently average the contributions and obtain a value of $1\,\%$. The uncertainty on the calibration (9.6\%) includes the uncertainty on the correction of $N_{B\overline{B}}$ (1.4\%) and the uncertainty on $\mathcal{B}\left(B\to D \pi\right)$. Based on studies using dedicated control samples, we assign $\unit[2.0]{\%}$, $\unit[4.0]{\%}$, and $\unit[2.2]{\%}$ for the uncertainties on PID efficiency, $\pi^0$ efficiency and \ks efficiency, respectively. The systematic uncertainty is included by convolving the likelihood function with a Gaussian with zero mean and a width equal to the square root of the quadratic sum of the additive and multiplicative error. The additive uncertainty is defined as the uncertainty on the signal yield, and contributions are summarized in \cref{tab:systematics}. A comparison of our results with previous ones is presented in \cref{fig:final-plot}.

\begin{table*}
  \centering
    \caption{Additive systematic uncertainties. All values are event yields.}
    \input{additive_systematics}
    \label{tab:systematics}
\end{table*}

The systematic uncertainties are evaluated using independent samples of MC and data control samples for charged and neutral modes. They can therefore be considered uncorrelated. Thus, we combine charged and neutral modes by adding the negative log likelihoods. We scale the branching fraction of the neutral modes by a factor of $\tau_{B^+}/\tau_{B^0}$ since the lifetime difference is the only factor distinguishing charged from neutral \chan{0} decays in the SM. We subsequently repeat the calculation of the limit and obtain the following values at $90\,\%$ C.L.:
\begin{align}
\mathscr{B}(B\to K\nu\overline{\nu}) &< 1.6\times 10^{-5},\notag\\
\mathscr{B}(B\to K^\ast\nu\overline{\nu}) &< 2.7\times 10^{-5},\notag\\
\mathscr{B}(B\to \pi\nu\overline{\nu}) &< 0.8\times 10^{-5},\notag\\
 \mathscr{B}(B\to \rho\nu\overline{\nu}) &< 2.8\times 10^{-5}.\notag
\end{align}

Based on the values and theoretical uncertainties from Ref.~\cite{Buras_knunu_14}, we also give a limit on the ratios between the measured branching fractions of $B\to K\nu\overline{\nu}$ and of $B\to K^\ast\nu\overline{\nu}$ and the respective SM prediction $\mathcal{R}_{K^{\ast}}$. We obtain values of $\mathcal{R}_K < 3.9$ and $\mathcal{R}_{K^\ast} < 2.7$, respectively, where we included the theoretical uncertainty. Both values are quoted at $\unit[90]{\%}$ C.L.

In summary, we report the results of a search for eight different $B$ decay channels with a pair of neutrinos in the final state, where the second $B$ is reconstructed in one of 108 semileptonic decay channels. No significant signal is observed and limits are set on the respective branching fractions at a confidence level of $\unit[90]{\%}$. The limits on the branching fraction for the \chan{310}\!\!, \chan{313}\!\!, \chan{211}\!\!, \chan{111}\!\!, \mbox{\chan{213}\!\!,} and \chan{113} channels are the most stringent to date. Although our analysis yields important improvements, none of these limits excludes SM predictions and all of them leave room for contributions from new physics.\\

We thank the KEKB group for the excellent operation of the
accelerator; the KEK cryogenics group for the efficient
operation of the solenoid; and the KEK computer group,
the National Institute of Informatics, and the 
PNNL/EMSL computing group for valuable computing
and SINET5 network support.  We acknowledge support from
the Ministry of Education, Culture, Sports, Science, and
Technology (MEXT) of Japan, the Japan Society for the 
Promotion of Science (JSPS), and the Tau-Lepton Physics 
Research Center of Nagoya University; 
the Australian Research Council;
Austrian Science Fund under Grant No.~P 26794-N20;
the National Natural Science Foundation of China under Contracts 
No.~10575109, No.~10775142, No.~10875115, No.~11175187, No.~11475187, 
No.~11521505 and No.~11575017;
the Chinese Academy of Science Center for Excellence in Particle Physics; 
the Ministry of Education, Youth and Sports of the Czech
Republic under Contract No.~LTT17020;
the Carl Zeiss Foundation, the Deutsche Forschungsgemeinschaft, the
Excellence Cluster Universe, and the VolkswagenStiftung;
the Department of Science and Technology of India; 
the Istituto Nazionale di Fisica Nucleare of Italy; 
the WCU program of the Ministry of Education, National Research Foundation (NRF)
of Korea Grants No.~2011-0029457, No.~2012-0008143,
No.~2014R1A2A2A01005286,
No.~2014R1A2A2A01002734, No.~2015R1A2A2A01003280,
No.~2015H1A2A1033649, No.~2016R1D1A1B01010135, No.~2016K1A3A7A09005603, No.~2016K1A3A7A09005604, No.~2016R1D1A1B02012900,
No.~2016K1A3A7A09005606, No.~NRF-2013K1A3A7A06056592;
the Brain Korea 21-Plus program, Radiation Science Research Institute, Foreign Large-size Research Facility Application Supporting project and the Global Science Experimental Data Hub Center of the Korea Institute of Science and Technology Information;
the Polish Ministry of Science and Higher Education and 
the National Science Center;
the Ministry of Education and Science of the Russian Federation and
the Russian Foundation for Basic Research;
the Slovenian Research Agency;
Ikerbasque, Basque Foundation for Science and
MINECO (Juan de la Cierva), Spain;
the Swiss National Science Foundation; 
the Ministry of Education and the Ministry of Science and Technology of Taiwan;
and the U.S.\ Department of Energy and the National Science Foundation.

\end{document}

%% file: variable_definitions.tex
\newcommand{\dl}{\ensuremath{D^{(\ast)}l}\:}
\newcommand{\hsig}{\ensuremath{h_{\text{sig}}}\:}
\newcommand{\EEcl}{\ensuremath{E_{\text{ECL}}}\:}
\newcommand{\tagout}{\ensuremath{\mathscr{N}_{\text{tag}}}\:}

\newcommand{\csout}{\ensuremath{\mathscr{N}_{\text{CS}}}\:}
\newcommand{\pcms}{\ensuremath{p_{\text{cms}}}\:}
\newcommand{\emiss}{\ensuremath{E_{\text{miss}}}\:}

\newcommand{\ctbdl}{\ensuremath{\cos{{\theta}_{B,\:\dl}}}\:}

\newcommand{\taglpcms}{\ensuremath{p_{l_{\text{tag}}}}\:}

\newcommand{\finalout}{\ensuremath{\mathscr{N}_{\mathrm{sel}}}\:}

%% file: owncommands.tex
\newcommand{\ks}{\ensuremath{K_{\text{S}}^0}\:}
\newcommand{\kstarP}{\ensuremath{{K^\ast}^+}}
\newcommand{\kstarO}{\ensuremath{{K^\ast}^0}}
\newcommand{\klong}{\ensuremath{K_{\text{L}}^0}}

\newlength{\pluswidth}
\newlength{\zeroheight}
\newcommand{\mychan}{\ensuremath{B\to h\nu\bar{\nu}}}
\newcommand{\dstarlnu}{\ensuremath{B\to D^{\ast}l\nu}}

\newcommand{\chan}[1]{
  \IfEqCase{#1}{
    {321}{\mbox{\ensuremath{B^+\to K^+\nu\bar{\nu}}}}%
    {310}{\mbox{\ensuremath{B^{0}_{\phantom{+}}\to \ks\nu\bar{\nu}}}}%
    {323}{\mbox{\ensuremath{B^+\to \kstarP\nu\bar{\nu}}}}%
    {3321}{\mbox{\ensuremath{B^+\to \kstarP\left(\to K^+\pi^0\right)\nu\bar{\nu}}}}%
    {3310}{\mbox{\ensuremath{B^+\to \kstarP\left(\to \ks\pi^+\right)\nu\bar{\nu}}}}%
    {313}{\mbox{\ensuremath{B^{0}_{\phantom{+}}\to \kstarO\nu\bar{\nu}}}}%
    {211}{\mbox{\ensuremath{B^+\to \pi^+\nu\bar{\nu}}}}%
    {111}{\mbox{\ensuremath{B^{0}_{\phantom{+}}\to \pi^0\nu\bar{\nu}}}}%
    {213}{\mbox{\ensuremath{B^+\to \rho^+\nu\bar{\nu}}}}%
    {113}{\mbox{\ensuremath{B^{0}_{\phantom{+}}\to \rho^0\nu\bar{\nu}}}}%
}[\mychan]
}

\newcommand{\shortchan}[1]{
  \IfEqCase{#1}{
    {321}{\mbox{\ensuremath{K^+\nu\bar{\nu}}}}%
    {310}{\mbox{\ensuremath{\ks\nu\bar{\nu}}}}%
    {323}{\mbox{\ensuremath{\kstarP\nu\bar{\nu}}}}%
    {3321}{\mbox{\ensuremath{\kstarP\left(\K^+\pi^0\right)\nu\bar{\nu}}}}%
    {3310}{\mbox{\ensuremath{\kstarP\left(\to \ks\pi^+\right)\nu\bar{\nu}}}}%
    {313}{\mbox{\ensuremath{\kstarO\nu\bar{\nu}}}}%
    {211}{\mbox{\ensuremath{\pi^+\nu\bar{\nu}}}}%
    {111}{\mbox{\ensuremath{\pi^0\nu\bar{\nu}}}}%
    {213}{\mbox{\ensuremath{\rho^+\nu\bar{\nu}}}}%
    {113}{\mbox{\ensuremath{\rho^0\nu\bar{\nu}}}}%
}[\mbox{\ensuremath{h\nu\bar{\nu}}}]
}

\newcommand{\boldchan}[1]{
  \IfEqCase{#1}{
    {321}{\mbox{\ensuremath{\boldsymbol{B^{+}\to K^{+}\nu\bar{\nu}}}}}%
    {310}{\mbox{\ensuremath{\boldsymbol{B^{0}_{\phantom{+}}\to \ks\nu\bar{\nu}}}}}%
    {323}{\mbox{\ensuremath{\boldsymbol{B^{+}\to \kstarP\nu\bar{\nu}}}}}%
    {3321}{\mbox{\ensuremath{\boldsymbol{B^{+}\to \kstarP\left(\to K^+\pi^0\right)\nu\bar{\nu}}}}}%
    {3310}{\mbox{\ensuremath{\boldsymbol{B^{+}\to \kstarP\left(\to \ks\pi^+\right)\nu\bar{\nu}}}}}%
    {313}{\mbox{\ensuremath{\boldsymbol{B^{0}_{\phantom{+}}\to \kstarO\nu\bar{\nu}}}}}%
    {211}{\mbox{\ensuremath{\boldsymbol{B^{+}\to \pi^{+}\nu\bar{\nu}}}}}%
    {111}{\mbox{\ensuremath{\boldsymbol{B^{0}_{\phantom{+}}\to \pi^0\nu\bar{\nu}}}}}%
    {213}{\mbox{\ensuremath{\boldsymbol{B^{+}\to \rho^{+}\nu\bar{\nu}}}}}%
    {113}{\mbox{\ensuremath{\boldsymbol{B^{0}_{\phantom{+}}\to \rho^0\nu\bar{\nu}}}}}%
}[\mychan]
}

\newcommand{\ctlchan}[1]{
  \IfEqCase{#1} {
    {411}{\mbox{\ensuremath{B^{\parbox{\widthof{+}}{0}}\to D^-\pi^+}\:}}%
    {413}{\mbox{\ensuremath{B^{\parbox{\widthof{+}}{0}}\to D^{\ast -}l^+\nu_l}\:}}%
    {421}{\mbox{\ensuremath{B^+\to \bar{D}^0\pi^+}\:}}%
    {423}{\mbox{\ensuremath{B^+\to \bar{D}^{\ast 0}l^+\nu_l}\:}}%
    {4110}{\mbox{\ensuremath{\boldsymbol{{B^{\parbox{\widthof{+}}{0}}}\to D^-\pi^{+}}}\:}}%
    {4130}{\mbox{\ensuremath{\boldsymbol{{B^{\parbox{\widthof{+}}{0}}}\to D^{\ast -}l^{+}\nu_{l}}}\:}}%
    {4210}{\mbox{\ensuremath{\boldsymbol{{B^+}\to \bar{D}^{0}\pi^{+}}}\:}}%
    {4230}{\mbox{\ensuremath{\boldsymbol{{B^+}\to \bar{D}^{\ast 0}{l^+}\nu_{l}}}\:}}%
}[$\dstarlnu$]
}

\newcommand{\YfourS}{\mbox{\ensuremath{\Upsilon\!\left(\mathrm{4S}\right)}\:}}
\newcommand{\btag}{\ensuremath{B_{\text{tag}}}\:}

\newcommand{\pid}[2][]{%
  \ifthenelse{\equal{#1}{}}{\ensuremath{\mathcal{P}_{#2}}}{\ensuremath{\mathcal{P}_{{#1} \!/ {#2}}}}
}


%% file: author.tex
\noaffiliation
\affiliation{University of the Basque Country UPV/EHU, 48080 Bilbao}
\affiliation{Beihang University, Beijing 100191}
\affiliation{University of Bonn, 53115 Bonn}
\affiliation{Budker Institute of Nuclear Physics SB RAS, Novosibirsk 630090}
\affiliation{Faculty of Mathematics and Physics, Charles University, 121 16 Prague}
\affiliation{Chonnam National University, Kwangju 660-701}
\affiliation{University of Cincinnati, Cincinnati, Ohio 45221}
\affiliation{Deutsches Elektronen--Synchrotron, 22607 Hamburg}
\affiliation{University of Florida, Gainesville, Florida 32611}
\affiliation{Justus-Liebig-Universit\"at Gie\ss{}en, 35392 Gie\ss{}en}
\affiliation{Gifu University, Gifu 501-1193}
\affiliation{SOKENDAI (The Graduate University for Advanced Studies), Hayama 240-0193}
\affiliation{Hanyang University, Seoul 133-791}
\affiliation{University of Hawaii, Honolulu, Hawaii 96822}
\affiliation{High Energy Accelerator Research Organization (KEK), Tsukuba 305-0801}
\affiliation{J-PARC Branch, KEK Theory Center, High Energy Accelerator Research Organization (KEK), Tsukuba 305-0801}
\affiliation{IKERBASQUE, Basque Foundation for Science, 48013 Bilbao}
\affiliation{Indian Institute of Technology Bhubaneswar, Satya Nagar 751007}
\affiliation{Indian Institute of Technology Guwahati, Assam 781039}
\affiliation{Indian Institute of Technology Madras, Chennai 600036}
\affiliation{Indiana University, Bloomington, Indiana 47408}
\affiliation{Institute of High Energy Physics, Chinese Academy of Sciences, Beijing 100049}
\affiliation{Institute of High Energy Physics, Vienna 1050}
\affiliation{INFN - Sezione di Torino, 10125 Torino}
\affiliation{J. Stefan Institute, 1000 Ljubljana}
\affiliation{Kanagawa University, Yokohama 221-8686}
\affiliation{Institut f\"ur Experimentelle Kernphysik, Karlsruher Institut f\"ur Technologie, 76131 Karlsruhe}
\affiliation{Kennesaw State University, Kennesaw, Georgia 30144}
\affiliation{King Abdulaziz City for Science and Technology, Riyadh 11442}
\affiliation{Department of Physics, Faculty of Science, King Abdulaziz University, Jeddah 21589}
\affiliation{Korea Institute of Science and Technology Information, Daejeon 305-806}
\affiliation{Korea University, Seoul 136-713}
\affiliation{Kyungpook National University, Daegu 702-701}
\affiliation{\'Ecole Polytechnique F\'ed\'erale de Lausanne (EPFL), Lausanne 1015}
\affiliation{P.N. Lebedev Physical Institute of the Russian Academy of Sciences, Moscow 119991}
\affiliation{Faculty of Mathematics and Physics, University of Ljubljana, 1000 Ljubljana}
\affiliation{Ludwig Maximilians University, 80539 Munich}
\affiliation{Luther College, Decorah, Iowa 52101}
\affiliation{University of Maribor, 2000 Maribor}
\affiliation{Max-Planck-Institut f\"ur Physik, 80805 M\"unchen}
\affiliation{School of Physics, University of Melbourne, Victoria 3010}
\affiliation{University of Miyazaki, Miyazaki 889-2192}
\affiliation{Moscow Physical Engineering Institute, Moscow 115409}
\affiliation{Moscow Institute of Physics and Technology, Moscow Region 141700}
\affiliation{Graduate School of Science, Nagoya University, Nagoya 464-8602}
\affiliation{Kobayashi-Maskawa Institute, Nagoya University, Nagoya 464-8602}
\affiliation{Nara Women's University, Nara 630-8506}
\affiliation{National Central University, Chung-li 32054}
\affiliation{National United University, Miao Li 36003}
\affiliation{Department of Physics, National Taiwan University, Taipei 10617}
\affiliation{H. Niewodniczanski Institute of Nuclear Physics, Krakow 31-342}
\affiliation{Nippon Dental University, Niigata 951-8580}
\affiliation{Niigata University, Niigata 950-2181}
\affiliation{Novosibirsk State University, Novosibirsk 630090}
\affiliation{Osaka City University, Osaka 558-8585}
\affiliation{Pacific Northwest National Laboratory, Richland, Washington 99352}
\affiliation{University of Pittsburgh, Pittsburgh, Pennsylvania 15260}
\affiliation{Theoretical Research Division, Nishina Center, RIKEN, Saitama 351-0198}
\affiliation{University of Science and Technology of China, Hefei 230026}
\affiliation{Showa Pharmaceutical University, Tokyo 194-8543}
\affiliation{Soongsil University, Seoul 156-743}
\affiliation{Sungkyunkwan University, Suwon 440-746}
\affiliation{School of Physics, University of Sydney, New South Wales 2006}
\affiliation{Department of Physics, Faculty of Science, University of Tabuk, Tabuk 71451}
\affiliation{Tata Institute of Fundamental Research, Mumbai 400005}
\affiliation{Excellence Cluster Universe, Technische Universit\"at M\"unchen, 85748 Garching}
\affiliation{Department of Physics, Technische Universit\"at M\"unchen, 85748 Garching}
\affiliation{Toho University, Funabashi 274-8510}
\affiliation{Department of Physics, Tohoku University, Sendai 980-8578}
\affiliation{Earthquake Research Institute, University of Tokyo, Tokyo 113-0032}
\affiliation{Department of Physics, University of Tokyo, Tokyo 113-0033}
\affiliation{Tokyo Institute of Technology, Tokyo 152-8550}
\affiliation{Tokyo Metropolitan University, Tokyo 192-0397}
\affiliation{University of Torino, 10124 Torino}
\affiliation{Virginia Polytechnic Institute and State University, Blacksburg, Virginia 24061}
\affiliation{Wayne State University, Detroit, Michigan 48202}
\affiliation{Yamagata University, Yamagata 990-8560}
\affiliation{Yonsei University, Seoul 120-749}

   \author{J.~Grygier}\affiliation{Institut f\"ur Experimentelle Kernphysik, Karlsruher Institut f\"ur Technologie, 76131 Karlsruhe} 
   \author{P.~Goldenzweig}\affiliation{Institut f\"ur Experimentelle Kernphysik, Karlsruher Institut f\"ur Technologie, 76131 Karlsruhe} 
   \author{M.~Heck}\affiliation{Institut f\"ur Experimentelle Kernphysik, Karlsruher Institut f\"ur Technologie, 76131 Karlsruhe} 

  \author{I.~Adachi}\affiliation{High Energy Accelerator Research Organization (KEK), Tsukuba 305-0801}\affiliation{SOKENDAI (The Graduate University for Advanced Studies), Hayama 240-0193} 
  \author{H.~Aihara}\affiliation{Department of Physics, University of Tokyo, Tokyo 113-0033} 
  \author{S.~Al~Said}\affiliation{Department of Physics, Faculty of Science, University of Tabuk, Tabuk 71451}\affiliation{Department of Physics, Faculty of Science, King Abdulaziz University, Jeddah 21589} 
  \author{D.~M.~Asner}\affiliation{Pacific Northwest National Laboratory, Richland, Washington 99352} 
  \author{T.~Aushev}\affiliation{Moscow Institute of Physics and Technology, Moscow Region 141700} 
  \author{R.~Ayad}\affiliation{Department of Physics, Faculty of Science, University of Tabuk, Tabuk 71451} 
  \author{T.~Aziz}\affiliation{Tata Institute of Fundamental Research, Mumbai 400005} 
% ...
\collaboration{The Belle Collaboration}

%% file: observed_events.tex
\begin{tabular*}{\linewidth}{@{\extracolsep{\fill}}lcrcrclcr}
  \toprule
  \toprule
  Channel&\multicolumn{7}{c}{Observed signal yield}&Significance\\
  \midrule
  \shortchan{321}&\phantom{mmm}&$17.7$&$\hspace{-0.75em}\pm\hspace{-0.75em}$&$9.1$&$\hspace{-0.75em}\pm\hspace{-0.75em}$&$3.4$&\phantom{mm}&$1.9\,\sigma$\\
  \shortchan{310}&\phantom{mmm}&$0.6$&$\hspace{-0.75em}\pm\hspace{-0.75em}$&$4.2$&$\hspace{-0.75em}\pm\hspace{-0.75em}$&$1.4$&\phantom{mm}&$0.0\,\sigma$\\
  \shortchan{323}&\phantom{mmm}&$16.2$&$\hspace{-0.75em}\pm\hspace{-0.75em}$&$7.4$&$\hspace{-0.75em}\pm\hspace{-0.75em}$&$1.8$&\phantom{mm}&$2.3\,\sigma$\\
  \shortchan{313}&\phantom{mmm}&$-2.0$&$\hspace{-0.75em}\pm\hspace{-0.75em}$&$3.6$&$\hspace{-0.75em}\pm\hspace{-0.75em}$&$1.8$&\phantom{mm}&$0.0\,\sigma$\\
  \shortchan{211}&\phantom{mmm}&$5.6$&$\hspace{-0.75em}\pm\hspace{-0.75em}$&$15.1$&$\hspace{-0.75em}\pm\hspace{-0.75em}$&$5.9$&\phantom{mm}&$0.0\,\sigma$\\
  \shortchan{111}&\phantom{mmm}&$0.2$&$\hspace{-0.75em}\pm\hspace{-0.75em}$&$5.6$&$\hspace{-0.75em}\pm\hspace{-0.75em}$&$1.6$&\phantom{mm}&$0.0\,\sigma$\\
  \shortchan{213}&\phantom{mmm}&$6.2$&$\hspace{-0.75em}\pm\hspace{-0.75em}$&$12.3$&$\hspace{-0.75em}\pm\hspace{-0.75em}$&$2.4$&\phantom{mm}&$0.3\,\sigma$\\
  \shortchan{113}&\phantom{mmm}&$11.9$&$\hspace{-0.75em}\pm\hspace{-0.75em}$&$9.0$&$\hspace{-0.75em}\pm\hspace{-0.75em}$&$3.6$&\phantom{mmm}&$1.2\,\sigma$\\
 \bottomrule
  \bottomrule
\end{tabular*}

%% file: measured_br_limits_result.tex
\begin{tabular*}{\linewidth}{@{\extracolsep{\fill}}lcrr}
\toprule
\toprule
Channel&Efficiency&Expected limit&Observed limit\\
\midrule
$\shortchan{321}$&$2.16\times 10^{-3}$&$0.8\times10^{-5}$&$1.9\times10^{-5}$\\
$\shortchan{310}$&$0.91\times 10^{-3}$&$1.2\times10^{-5}$&$1.3\times10^{-5}$\\
$\shortchan{323}$&$0.57\times 10^{-3}$&$2.4\times10^{-5}$&$6.1\times10^{-5}$\\
$\shortchan{313}$&$0.51\times 10^{-3}$&$2.4\times10^{-5}$&$1.8\times10^{-5}$\\
$\shortchan{211}$&$2.92\times 10^{-3}$&$1.3\times10^{-5}$&$1.4\times10^{-5}$\\
$\shortchan{111}$&$1.42\times 10^{-3}$&$1.0\times10^{-5}$&$0.9\times10^{-5}$\\
$\shortchan{213}$&$1.11\times 10^{-3}$&$2.5\times10^{-5}$&$3.0\times10^{-5}$\\
$\shortchan{113}$&$0.82\times 10^{-3}$&$2.2\times10^{-5}$&$4.0\times10^{-5}$\\
\bottomrule
\bottomrule
\end{tabular*}

%% file: additive_systematics.tex
\begin{tabular*}{\linewidth}{@{\extracolsep{\fill}}lrrrrrrrr}
  \toprule
  \toprule
  &\hspace{0.5em}\shortchan{321}&\hspace{0.5em}\shortchan{310}&\hspace{0.5em}\shortchan{323}&\hspace{0.5em}\shortchan{313}&\hspace{0.5em}\shortchan{211}&\hspace{0.5em}\shortchan{111}&\hspace{0.5em}\shortchan{213}&\hspace{0.5em}\shortchan{113}\\
  \midrule
  \klong~veto&0.2&0.2&0.1&0.2&0.6&0.4&0.6&0.0\\
  Fixed fractions&0.4&0.3&0.1&0.2&1.3&0.1&0.1&1.0\\
  Continuum scaling&2.0&0.0&0.0&0.0&3.1&0.0&0.0&0.0\\
  Tag efficiency correction&0.5&0.2&0.1&0.1&1.9&0.1&0.2&0.5\\
  Shape uncertainty&2.6&1.3&1.8&1.7&4.5&1.5&2.3&3.4\\
  Fit bias&0.2&0.1&0.2&0.1&0.2&0.1&0.2&0.2\\
  \midrule
  Total&3.4&1.4&1.8&1.8&5.9&1.6&2.4&3.6\\
  \bottomrule
  \bottomrule
\end{tabular*}

%% file: hnunu_paper.bbl
\begin{thebibliography}{99}
\bibitem{charge_conj}
Throughout this letter, $h$ refers to one of the following charmless states: $K^+$, $\ks$, $K^{\ast +}$, $K^{\ast 0}$, $\pi^+$, $\pi^0$, $\rho^+$, $\rho^0$. Charge-conjugate channels are implied throughout this paper unless explicitly stated otherwise.
\bibitem{ctbdl}
The cosine of the angle between the $B$ and the \dl\: system is defined as $\ctbdl=\left(2E_{B}E_{\dl}-M_B^2c^4-M_{\dl}^2c^4\right)/\left(2\left|\mathbf{p}_B\right|\left|\mathbf{p}_{\dl}\right|\!c^2\right)$ where $M$, $E$, and $\mathbf{p}$ are the invariant mass, energy, and momentum of the $B$ (\dl) system.
\bibitem{CL}
As we use a Bayesian method, this is formally a ``credibility level.'' However, we use ``confidence level'' here following common convention.
\end{thebibliography}
